\documentclass[
  ,final            
  ]
  {aipproc}

\usepackage{bm}

\layoutstyle{8x11single}

\def\ve{\varepsilon}

\begin{document}

\title{Event-by-event hydrodynamics for heavy-ion collisions}


\author{Zhi Qiu}{
  address={Department of Physics, The Ohio State University, Columbus, Ohio 43026, USA}
}

\author{Ulrich Heinz}{
  address={Department of Physics, The Ohio State University, Columbus, Ohio 43026, USA}
}

\begin{abstract}

We compare $v_2/\ve_2$ and $v_3/\ve_3$ from single-shot and 
event-by-event (2+1)-dimensional hydrodynamic calculations and discuss 
the validity of using single-shot calculations as substitutes for event-by-event 
calculations. Further we present a proof-of-concept calculation demonstrating 
that $v_2$ and $v_3$ together can be used to strongly reduce initial condition 
ambiguities.

\end{abstract}

\maketitle


\section{Introduction}

The Quark-Gluon-Plasma (QGP) created in heavy-ion collisions has been under 
intense study. In particular, it has been shown to exhibit almost perfect liquid collective 
behaviour. The extraction of one of its transport coefficients, the specific shear viscosity 
$\eta/s$, has recently become one of the hottest topics.

Shear viscosity reduces the conversion efficiency from initial geometry deformation to 
final flow anisotropies, and in \cite{Qiu:2011iv} we reported that the eccentricity-scaled 
second and third order harmonic flow coefficients, $v_2/\ve_2$ and $v_3/\ve_3$, for 
unidentified charged hadrons are good choices to extract $\eta/s$. For $v_2/\ve_2$ 
this has already been done by several groups, but it was shown that this leaves a large 
uncertainty for $\eta/s$ due to ambiguities in the initial fireball deformation (see e.g. 
\cite{Luzum:2008cw,Song:2010mg}). We here show that a simultaneous analysis of 
$v_2/\ve_2$ and $v_3/\ve_3$ can resolve this ambiguity. The authors of  
\cite{Luzum:2008cw,Song:2010mg} use a "single-shot" hydrodynamic approach
which evolves a smooth initial profile obtained by averaging over an ensemble of 
fluctuating bumpy initial conditions. We here address the question if it matters whether 
one instead follows nature's example and evolves each bumpy initial condition 
separately ("event-by-event hydrodynamics"), averaging over the fluctuating event 
ensemble only at the end. Due to the high numerical cost of the event-by-event approach 
on the one hand and the strong sensitivity of $\eta/s$ on $v_2/\ve_2$ and $v_3/\ve_3$
on the other hand this issue is of high practical relevance.

\section{Single-shot vs. event-by-event hydrodynamic calculations}

We initialize the hydrodynamic simulations by generating fluctuating initial entropy density 
profiles from the Monte Carlo Glauber (MC-Glb.) and Monte Carlo KLN (MC-KLN) models 
for Au+Au collisions at $\sqrt{s}=200\,A$\,GeV (see \cite{Qiu:2011iv} for details).
%
\begin{figure}[ht]
  \includegraphics[width=0.84\textwidth]{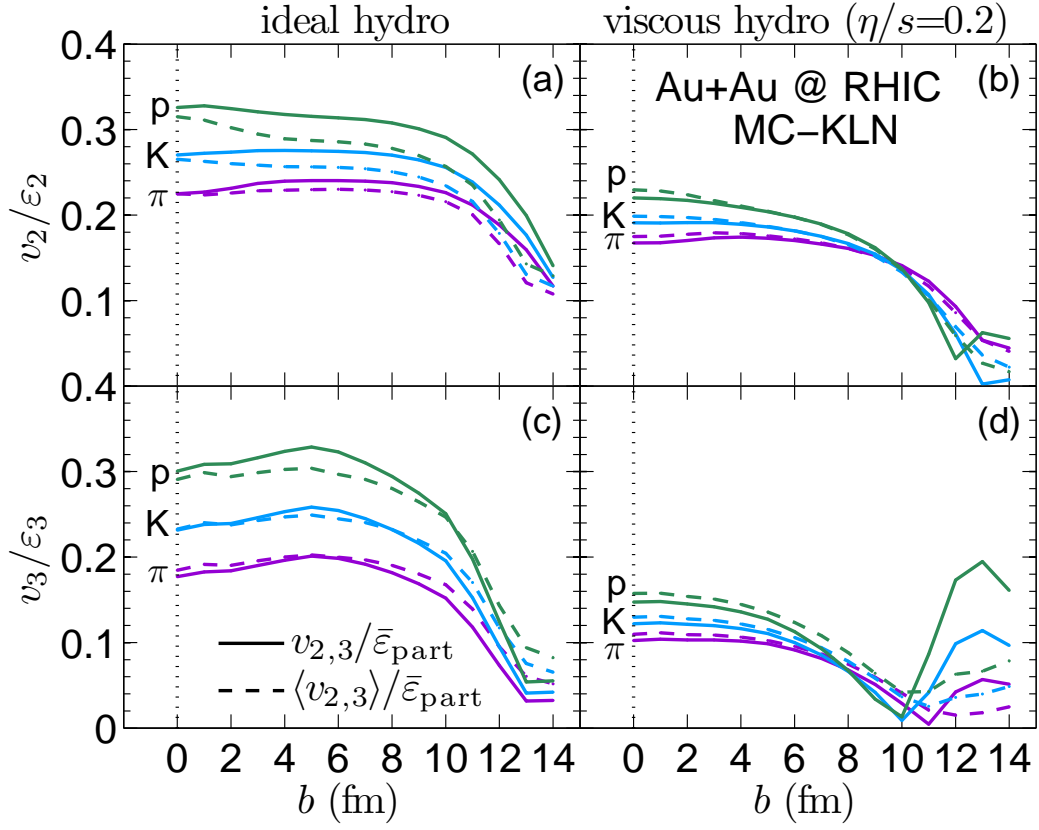}
  \caption{Eccentricity-scaled elliptic ($v_2/\ve_2$, top) and triangular flow 
  ($v_3/\ve_3$, bottom) as a function of impact parameter, for thermal pions, 
  kaons, and protons from 200\,$A$\,GeV Au+Au collisions. Solid and dashed 
  lines show results from single-shot and event-by-event hydrodynamics, 
  respectively, for MC-KLN initial conditions. The left panels (a,c) 
  assume ideal \cite{Qiu:2011iv}, the right panels (b,d) viscous fluid dynamic 
  evolution with $\eta/s=0.2$.
  \label{fig:1}}
\end{figure}
%
In Fig.~\ref{fig:1} we compare the eccentricity-scaled elliptic and triangular flow 
coefficients for thermal pions, kaons and protons (i.e. without resonance decay 
contributions) from single-shot and event-by-event hydrodynamics, for ideal and
viscous fluids. For the single-shot simulations, we average over the event ensemble
in the participant plane, i.e. after centering and rotating each event by the participant 
plane angle $\psi_{2,3}^\mathrm{PP}(e)$ between the short axis of the second 
resp. third order harmonic component of its energy density profile $e(x,y)$ and the 
impact parameter $\bm{b}$ \cite{Qiu:2011iv}, before starting the evolution. 

For an ideal fluid (Fig.~\ref{fig:1}(a,c)) event-by-event hydrodynamics produces
significantly less elliptic and slightly less triangular flow than the equivalent
single-shot evolution. The difference in $v_2/\ve_2$ is smallest for pions 
(${\cal O}(5\%)$) but increases with hadron mass to about 10\% for protons. For 
$v_3/\ve_3$ the differences are smaller, but again increase with hadron mass.
In contrast, viscous hydrodynamics with $\eta/s=0.2$ produces almost the same
$v_2/\ve_2$ and $v_3/\ve_3$ for single-shot and event-by-event hydro,
irrespective of hadron mass; if anything, the flows are now a little larger when the
fluctuating events are evolved individually. Apparently, shear viscosity quickly 
damps the initial density fluctuations into something approaching the smooth 
ensemble-averaged profile before most of the flow develops. (The strange pattern
seen at very large impact parameters in Figs.~\ref{fig:1}(b,d) signals a jump of
the flow angle $\psi_{2,3}^\mathrm{EP}$ \cite{Qiu:2011iv} by $\pi/n$ ($n=2,3$.)

Fig.~\ref{fig:1} suggests that for a viscous fluid with sufficiently large viscosity
$\eta/s$ single-shot hydrodynamics can substitute for event-by-event evolution 
for the calculation of both $v_2/\ve_2$ and $v_3/\ve_3$ for unidentified and 
identified hadrons. For an ideal fluid this remains true for the triangular flow of
unidentified charged hadrons (which are mostly pions) but not for that of identified
heavy hadrons (e.g. protons), nor for the elliptic flow $v_2/\ve_2$. Until the QGP 
viscosity is known, it is therefore advisable to extract it from event-by-event hydrodynamic
simulations. If it turns out large enough, additional systematic studies can be done
using the more economic single-shot approach. 

 By comparing the left and right panels in Fig.~\ref{fig:1} we see that $\eta/s=0.2$ 
 suppresses $v_3/\ve_3$ much more strongly (by $\sim50\%$) than $v_2/\ve_2$ 
 (which is suppressed only by $\sim25\%$). Taken together, $v_2/\ve_2$ and $v_3/\ve_3$ 
 thus over-constrain $\eta/s$ for a given model of initial state eccentricities $\ve_n$.
 We will now show how this allows to distinguish experimentally between the 
 MC-Glauber and MC-KLN models.

\section{Reducing the initial condition model ambiguity}

We reported in \cite{Qiu:2011iv} that the MC-Glauber and MC-KLN models have similar 
$\ve_3$ but the MC-KLN model has $\sim20\%$ larger $\ve_2$, and that for ideal fluid
dynamics they give similar $v_2/\ve_2$ and $v_3/\ve_3$ ratios. Accordingly, for ideal
fluids the two models generate similar $v_3$ but different $v_2$. We expect the ratios 
$v_2/\ve_2$ and $v_3/\ve_3$ for these two models to remain similar even when adding
viscosity (a corresponding study is ongoing). Therefore, for any fixed $\eta/s$, they will 
generate similar $v_3$ but different $v_2$. Alternatively,  when using different $\eta/s$ 
for the two models so that they produce the same $v_2$, they will necessarily generate
different $v_3$. They will therefore not be able to simultaneously describe a given set
of experimental $v_2$ and $v_3$ data with the same medium properties.

To verify this statement quantitatively requires a proper event-by-event hydrodynmical 
calculation which is in progress. We here show instead results from a proof-of-concept 
calculation that supports our argument. We first generate a large set of deformed 
Gaussian initial conditions similar to the ones described in \cite{Alver:2010dn}, but having 
both non-zero $\ve_2$ and $\ve_3$, as calculated from the MC-Glauber and MC-KLN 
models for the 20-30\% centrality bin, with random relative orientation 
$\psi_3^\mathrm{PP}{-}\psi_2^\mathrm{PP}$ . We call these initial conditions "MC-Glauber-like" 
and "MC-KLN-like". 

\begin{figure}[h]
  \includegraphics[height=.28\textheight]{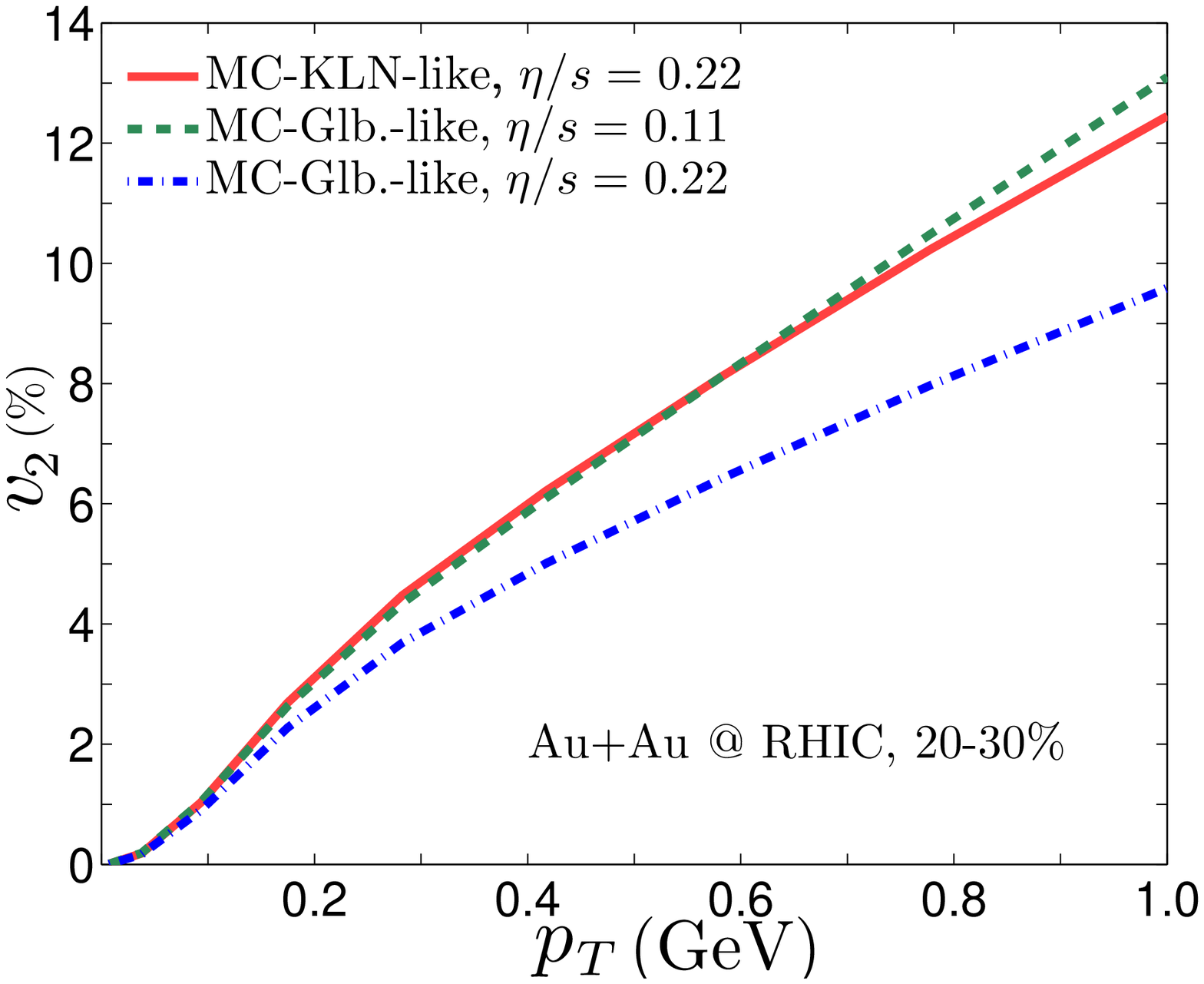}
  \includegraphics[height=.28\textheight]{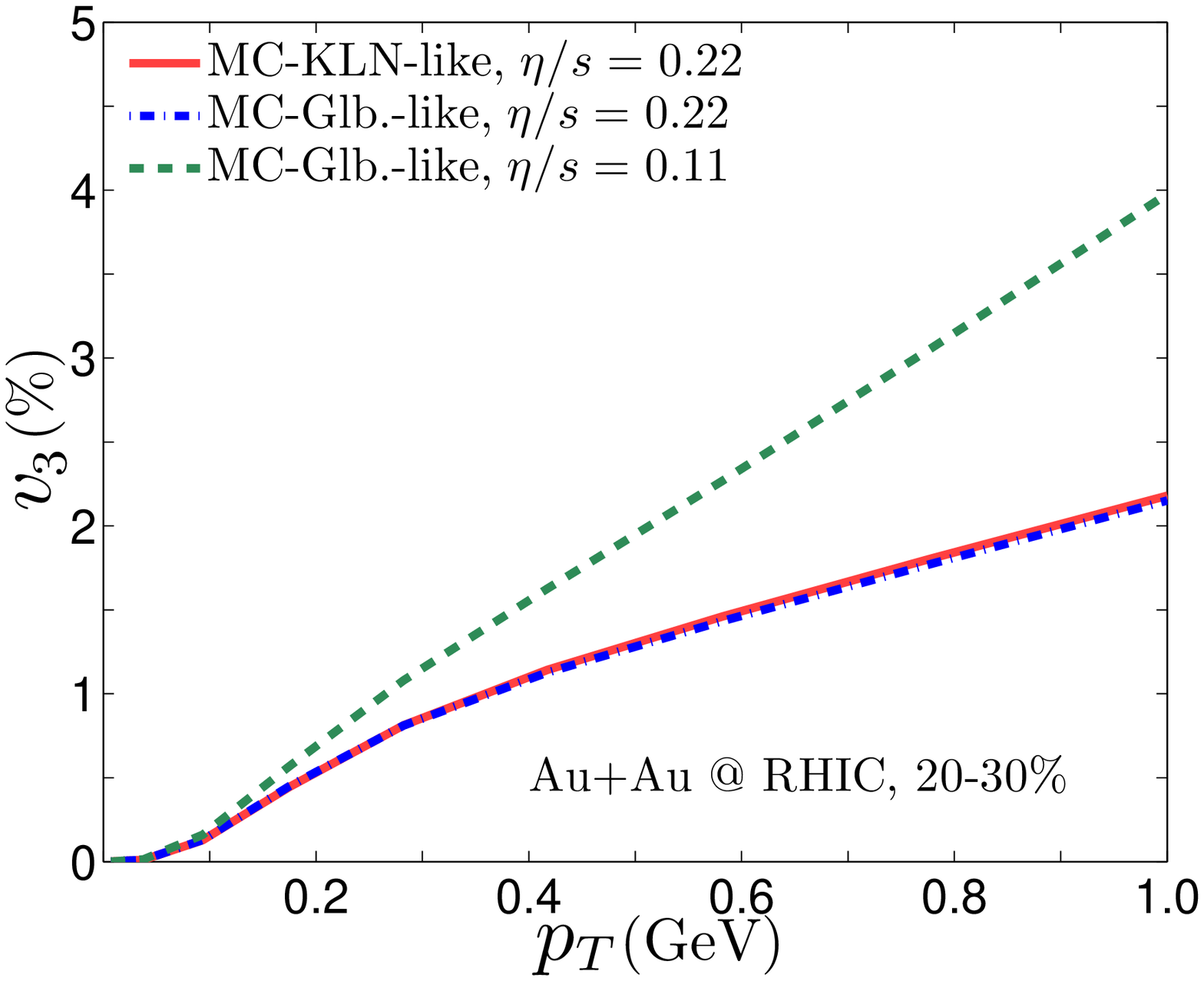}
  \caption{Differential $v_2(p_T)$ (left) and $v_3(p_T)$ (right) from viscous
  hydrodynamics using MC-Glauber-like and MC-KLN-like initial conditions and different
  values for $\eta/s$ (see text for discussion). 
  \label{fig:3}}
\end{figure}

Fig. \ref{fig:3} shows differential $v_{2,3}(p_T)$ curves resulting from the viscous 
hydrodynamic evolution of these initial conditions. The solid and dashed curves in the
left panel show that, in order to obtain the same $v_2(p_T)$ for MC-KLN-like and 
MC-Glauber like initial conditions, the fluid must be twice as viscous for the former than
for the latter. The right panel shows that, with $\eta/s$ chosen to produce the same 
$v_2$, MC-Glauber-like and MC-KLN-like initial conditions produce dramatically 
different $v_3$, with the one from MC-KLN-like initialization being much smaller.
Conversely, if $\eta/s$ is tuned to produce the same $v_3$, MC-Glauber-like and 
MC-KLN-like initial conditions require the same value of $\eta/s$ (solid and dash-dotted
lines in the right panel), which then leads to dramatically different $v_2$ values
for the different initial conditions (see corresponding lines in the left panel). These
conclusions agree qualitatively with corresponding statements made in 
Refs.~\cite{:2011vk,Adare:2011tg}.

\section{Summary}

We demonstrated that for sufficiently large viscosity ($\eta/s{\,>\,}{\cal O}(0.2)$) and 
limited precision requirements single-shot evolution of smooth averaged initial 
profiles can substitute for event-by-event evolution of fluctuating initial conditions, 
and that a simultaneous analysis of $v_2$ and $v_3$ overconstrains $\eta/s$ and
thus has the power to discriminate between initial state models. A precise extraction 
of $\eta/s$ without initial state ambiguity will, however, require event-by-event viscous 
hydrodynamical evolution of fluctuating initial states, coupled to a microscopic hadronic
cascade for the freeze-out stage \cite{Song:2010mg}, to calculate ensemble averages
of both $v_2$ and $v_3$.

\bigskip


\noindent
{\bf Acknowledgments:} This work was supported by the U.S.\ Department of Energy under grants 
No. \rm{DE-SC0004286} and (within the framework of the JET Collaboration) 
No. \rm{DE-SC0004104}.



\bibliographystyle{aipproc}   

\begin{thebibliography}{99}

\bibitem{Qiu:2011iv}
  Z.~Qiu and U.~Heinz,
  {\it Preprint arXiv:1104.0650 [nucl-th]}.

\bibitem{Luzum:2008cw}
  M.~Luzum and P.~Romatschke,
  Phys.\ Rev.\  C {\bf 78}, 034915 (2008)
  [Erratum {\it ibid.}\  C {\bf 79}, 039903 (2009)].
  
\bibitem{Song:2010mg}
  H.~Song, S.~A.~Bass, U.~Heinz, T.~Hirano and C.~Shen,
  Phys.\ Rev.\ Lett.\  {\bf 106}, 192301 (2011).

\bibitem{Alver:2010dn}
  B.~H.~Alver, C.~Gombeaud, M.~Luzum and J.~Y.~Ollitrault,
  Phys.\ Rev.\  C {\bf 82}, 034913 (2010).

\bibitem{:2011vk}
   K. Aamodt {\it et al.}  [ALICE Collaboration],
  Phys.\ Rev.\ Lett.\  {\bf 107}, 032301 (2011).

\bibitem{Adare:2011tg}
  A.~Adare {\it et al.}  [PHENIX Collaboration],
  {\it Preprint arXiv:1105.3928 [nucl-ex]}.

\end{thebibliography}

\end{document}